\documentstyle[multicol,aps,psfig]{revtex}
\begin{document}

\newcommand{\T}{\eta}           
\newcommand{\E}{E}              
\newcommand{\A}{A}              
\newcommand{\freevol}{\bar{v_{\rm f}}}  
\newcommand{\lta}{\stackrel{<}{\scriptstyle\sim}}  
\newcommand{\gta}{\stackrel{>}{\scriptstyle\sim}}  

\draft

\title{A phenomenological glass model for vibratory granular compaction}

\author{D. A. Head\cite{em1}}

\address{Department of Physics and Astronomy, JCMB King's Buildings,
University of Edinburgh, Edinburgh EH9 3JZ, United Kingdom}

\date{\today}

\maketitle

\begin{abstract}
A model for weakly excited granular media is derived by combining
the free volume argument of Nowak {\em et al.}
[Phys. Rev. E {\bf 57}, 1971 (1998)]
and the phenomenological model for supercooled liquids of Adam
and Gibbs [J. Chem. Phys. {\bf 43}, 139 (1965)].
This is made possible by relating the granular excitation parameter
$\Gamma$, defined as the peak acceleration of the driving pulse scaled
by gravity, to a temperature-like parameter~$\T(\Gamma)$.
The resulting master equation is formally identical to that of
Bouchaud's trap model for glasses [J. Phys. I {\bf 2}, 1705 (1992)].
Analytic and simulation results are shown to compare favourably with
a range of known experimental behaviour.
This includes the logarithmic densification and power spectrum
of fluctuations under constant~$\T$,
the annealing curve when $\T$ is varied cyclically in time,
and memory effects observed for a discontinuous shift in~$\T$.
Finally, we discuss the physical interpretation of the model parameters
and suggest further experiments for this class of systems.
\end{abstract}


\maketitle
\begin{multicols}{2}
\narrowtext

\section{Introduction}
\label{s:intro}

It is well known that, with the appropriate driving and boundary
conditions, granular matter can approximate each of
the three major states of matter: gas, liquid and solid~\cite{rev1,rev2}.
Conspicuous by its absence is a glass state; that is, a state where
the relaxation times far exceed the observational
timeframe~\cite{Nagelrev,Angellrev,Hammann,StAndrews,Nagata}.
However, it is becoming increasingly clear that the granular analogue
of glass has been found in a recent series of experiments
performed at the University of Chicago~\cite{exp1,exp2,exp3,exp4}.
They measured the density of a system that was weakly
perturbed or `tapped' by the application of a driving pulse to the container.
A first indication of glass-like relaxation processes
came from analysis of the density $\rho(t)$,
where $t$ is the number of times the system had been tapped,
which was found to increase only logarithmically slowly~\cite{exp1},

\begin{equation}
\rho(t)=\rho_{\rm f}-\frac{\Delta\rho}{1+B\ln(1+t/\tau)}\:\:.
\label{e:logrelax}
\end{equation}

\noindent{}The fitting parameters $\rho_{\rm f}$\,, $\Delta\rho$, $B$ and
$\tau$ are functions of the control parameter~$\Gamma$,
defined as the peak acceleration of the
driving pulse scaled by gravity, $\Gamma=a_{\rm max}/g$.
Subsequent experiments in which $\Gamma$ was varied during a run
also behaved in a similar manner to glasses
under a variable temperature~\cite{Hammann,StAndrews,Nagata,exp2,exp3,exp4},
suggesting a relationship between $\Gamma$ and some elusive
temperature-like quantity.

Theoretical attempts to understand the experiments have
ranged from the construction of toy microscopic models
to higher level, coarse grained descriptions
\cite{tetris_prl,tetris_full,tetris_aging,Mario_logdomain,Mario_FDT,Prados,parkinglot,Talbot,Gavrilov,BS,Sinai,Grinev,Linz,Barker}.
The general consensus has been that the slow relaxation is due to
frustrated dynamics resulting from excluded volume effects.
More insightful are the free volume arguments postulated
by the Chicago group~\cite{exp2} and Boutreux and de Gennes~\cite{deGennes},
which derive the logarithmic compaction with only a small number
of assumptions.
Provocatively, these assumptions are also key components in established
phenomenological models of glass-forming liquids,
namely those of Adam and Gibbs~\cite{AdamGibbs} and
Cohen and Turnbull~\cite{Cohen}, respectively.
This further suggests that the analogy with glasses is a valid one.
However, both of the granular free volume descriptions currently lack
any mention of the experimental control parameter
$\Gamma$ and hence must be regarded as incomplete.

In this paper we demonstrate how one of the free volume arguments,
namely that of the Chicago group, can be expanded into a full
model that incorporates~$\Gamma$.
This is made possible by
postulating a loose analogy between $\Gamma$ in the granular system
and temperature in supercooled liquids,
and then using this analogy to incorporate elements of the
Adam and Gibbs theory.
The result of this process is a master equation for weakly
excited granular media that is capable of reproducing a wide range
of known experimental behaviour.
The motivations behind this work are twofold.
Firstly, by focusing on only a small number of physical mechanisms,
the success of the model in emulating the experiments
indicates that the dominant mechanisms may have been correctly identified.
It is further hoped that this work may help to strengthen the relationship
between granular matter and glasses.
This second goal is easily achieved once we show that the
derived master equation is identical to that of a simple glass
model due to Bouchaud~\cite{trap_weak,trap_tree,trap_int,trap_jphysa}.

This paper is arranged as follows.
In Sec.~\ref{s:model} the Chicago group's free volume argument is summarised
and then expanded to a full model by importing elements of the
Adam and Gibbs theory.
The resulting master equation that describes the evolution of the
system in time is specified.
Numerical integration of this master equation,
plus analytical results wherever possible,
are compared to the experimental data in Sec.~\ref{s:exp}.
Of particular importance here is an explanation for the apparent
contradiction between the experiments and any model based
on the free volume approach,
concerning the supposed $\Gamma$--dependence of the
projected final density, $\rho_{\rm f}$ in~(\ref{e:logrelax}).
Further discussion on the physical interpretation of the model
parameters is given in Sec.~\ref{s:interp}, as well as
suggested ways in which the various assumptions behind the model
may be more rigorously checked.
Finally, we summarise our findings in Sec.~\ref{s:summ} and make
some tentative predictions for future experiments that may help to
further elucidate the relevant physical mechanisms in granular compaction.

\section{Description of the model}
\label{s:model}

The relationship between the Chicago group's free volume description
and the Adam and Gibbs theory is that they both regard the dominant
relaxation process to be the cooperative rearrangement of particles.
The correspondence between the two theories can be taken further
by postulating the existence of a temperature-like {\em noise}
parameter $\T(\Gamma)$ for weakly excited granular matter.
This procedure forms the basis of our work, and is described in full below.
For current purposes it is sufficient to provide a somewhat heuristic
description of the model; a fuller discussion of the various parameters
can be found in Sec.~\ref{s:interp}.

\subsection{First principles derivation}
\label{s:freevol}

The Chicago group's~\cite{exp2}, and also
Boutreux and de Gennes'~\cite{deGennes},
arguments employ the concept of the mean free volume per particle,
here denoted~$\bar{v_{\rm f}}$\,.
For a system of $N$ particles
occupying a total volume $V$, $\bar{v_{\rm f}}$ is defined as

\begin{equation}
\bar{v_{\rm f}}
=\frac{V-V_{\rm min}}{N}
=v_{\rm g}\left(\frac{1}{\rho}-\frac{1}{\rho_{\rm max}}\right)\:\:,
\label{e:vfbar}
\end{equation}

\noindent{}where $v_{\rm g}$ is the volume of a single particle and
$\rho$ is the volume fraction \mbox{$Nv_{\rm g}/V$.}
Units are chosen so that the density of a single grain is unity,
hence $\rho$ is also the density of the system.
Following~\cite{deGennes}, $\rho_{\rm max}\equiv Nv_{\rm g}/V_{\rm min}$
is identified
with the most compact state possible in a disordered system,
{\em i.e.} the random close-packing limit.
In what follows we shall fix $\rho_{\rm max}=0.64$,
believed to be the random close packed density for a system of
monodisperse spheres~\cite{rcp_anon}.

The Chicago group postulated that the compaction process is dominated
by the cooperative rearrangement of local domains of particles.
If $z$ is the number of particles in a region that
can rearrange independently of its environment, they argued
there is a lower cut-off

\begin{equation}
z\geq z^{*}=a\,\frac{v_{\rm g}}{\freevol}
\label{e:minsize}
\end{equation}

\noindent{}below which there is not enough free volume available
to allow reconfiguration.
Roughly speaking, $z^{*}$ is the number of particles that,
by adding up their individual free volumes, can make a single
`hole' big enough to allow exactly one particle to fit through.
The explicit dependence of $z^{*}$ on $\rho$ is found by
combining (\ref{e:vfbar}) and (\ref{e:minsize}),

\begin{equation}
z^{*} = a \left(
\frac{1}{\rho}-\frac{1}{\rho_{\rm max}}
\right)^{-1}\:\:.
\label{e:z_star}
\end{equation}

\noindent{}By assuming that the density increases at a rate proportional
to ${\rm e}^{-z^{*}}$, it is now possible to derive the logarithmic
compaction law $\rho(t)\sim -1/\ln(t)$~\cite{exp2}.

As mentioned in the introduction, the theory just outlined
is incomplete as it does not incorporate the experimental control
parameter~$\Gamma$.
In an attempt to resolve this deficiency, we observe that
a similar description for cooperative relaxation is also central
to the theory proposed by Adam and Gibbs for structural relaxation
in supercooled liquids~\cite{AdamGibbs}.
An intermediate stage of their calculations is of interest
here, namely that the relaxation rate $W$
can be expressed as a function of temperature $T$ as

\begin{equation}
W(T)\propto\exp\left(-\,\frac{z^{*}\Delta\E}{k_{\rm B}\,T}\right)\:\:,
\label{e:adamgibbs}
\end{equation}

\noindent{}where $\Delta\E$ is the free energy barrier per particle,
$k_{\rm B}$ is Boltzmann's constant and $z^{*}$ is again the smallest
number of particles that can rearrange independently of
their environment (which was ultimately related to the
configurational entropy).

The principle assumption behind our current work is that an
expression analogous to (\ref{e:adamgibbs}) also holds for weakly
excited granular media.
More precisely, we propose that a region with local density $\rho$
reconfigures at a rate

\begin{equation}
W(\rho,\Gamma)\propto
\exp\left(-\,\frac{z^{*}(\rho)\Delta\E}{\eta(\Gamma)}\right)\:\:,
\label{e:micrate}
\end{equation}

\noindent{}where $z^{*}$ is related to $\rho$ via~(\ref{e:z_star}).
$\Delta\E$ can be interpreted as a gravitational potential energy
barrier per particle,
and $\eta(\Gamma)$ gives some measure of the degree of excitation of
the system.
Note that although $\eta(\Gamma)$ plays the role of $k_{\rm B}\,T$,
we stop short of referring to it as a `granular temperature'
and instead regard it as a noise parameter
which is {\em defined} by~(\ref{e:micrate}),
with the only restriction that $\eta(\Gamma)$ should be a
monotonic increasing function of~$\Gamma$.
In what follows $\T(\Gamma)$ is essentially treated as a fitting parameter.
The physical meaning of $\eta(\Gamma)$ and $\Delta\E$
is discussed further in Sec.~\ref{s:interp}.

To fully specify the model, some rule is required that gives the
density of a region after it has reconfigured.
In general this will depend on its density before reconfiguration
as well as $\T(\Gamma)$, but for simplicity we shall ignore such
considerations here and simply assume that the density after
reconfiguration is given by the fixed
probability density function~$\mu'(\rho)$.
Specifically, $\mu'(\rho)\,{\rm d}\rho$ is the probability that
a region `falls' into a configuration with a density in the range
$[\rho,\rho+{\rm d}\rho)$.
The prior distribution $\mu'(\rho)$
(re-expressed in terms of the total energy barrier $E$ -- see below)
will play a central role in our model, although it shall
be demonstrated that, over timescales relevant to the
experiments, the model is essentially robust to the particular choice
of~$\mu'(\rho)$.
This is fortuitous, as the precise form of $\mu'(\rho)$
is unknown and we have instead considered
a range of plausible functional forms.

\subsection{Summary of the model}
\label{s:mod_summ}

Since the reconfiguration rate $W(\rho,\Gamma)$ given in (\ref{e:micrate})
depends on $\rho$ only via the total energy
barrier $\E=z^{*}(\rho)\Delta\E$,
it is convenient to now make the change of variables $\rho\rightarrow\E$,
where

\begin{mathletters}
\label{e:mapping}
\begin{eqnarray}
\E&=&z^{*}\Delta\E
=A\left(\frac{1}{\rho}-\frac{1}{\rho_{\rm max}}\right)^{-1}
\:\:,\label{e:efromrho}\\
A&=&a\,\Delta\E\:\:,
\label{e:covar}
\end{eqnarray}
\end{mathletters}

\noindent{}which is one-to-one and hence invertible for all 
$\rho\in[0,\rho_{\rm max})$ and $\E\in[0,\infty)$.
Thus, in what follows, the state of the system at any given time $t$
will in the first place be defined by the distribution of
energy barriers~$P(\E,t)$, and only then shall the mean density $\rho(t)$
be found by inverting the mapping~(\ref{e:mapping}) and averaging
over~$P(\E,t)$, {\em i.e.}

\begin{equation}
\rho(t)=\int_{0}^{\infty}\,
\frac{P(\E,t)}{\frac{A}{E}+\frac{1}{\displaystyle\rho_{\rm max}}}
\,{\rm d}\E\:\:.
\label{e:rhofrome}
\end{equation}

\noindent{}Note that, in principle, small values of $E$ should be disallowed
to reflect the fact that low density configurations are not mechanically
stable and will not arise.
For the sake of simplicity we choose to ignore this subtlety here.

The master equation for $P(\E,t)$ can be derived as follows.
The rate at which a region with a local barrier $\E(\rho)$
reconfigures is given by $\omega_{0}\,{\rm e}^{-\E/\T}$,
where the constant $\omega_{0}$ fixes the timescale.
After reconfiguring, the region falls into a state with a
new barrier $\E_{\rm new}$ with a probability~$\mu(\E_{\rm new})$,
where $\mu(\E)$ is just $\mu'(\rho)$ after the change of variables,
$\mu(\E){\rm d}\E=\mu'(\rho){\rm d}\rho$.
Assuming that the number of taps $t$ can be well approximated as
a continuous variable, $P(\E,t)$ evolves in time according to

\begin{mathletters}
\label{e:master}
\begin{eqnarray}
\frac{1}{\omega_{0}}\frac{\partial P(\E,t)}{\partial t}&=&
-\,{\rm e}^{-\E/\T}P(\E,t)+\omega(t)\,\mu(\E)\:\:,
\label{e:dpbydt}\\
\omega(t)&=&\int_{0}^{\infty}{\rm e}^{-\E/\T}P(\E,t)\,{\rm d}\E\:\:.
\label{e:omega_t}
\end{eqnarray} 
\end{mathletters}

\noindent{}The first and second terms on the right hand side of
(\ref{e:dpbydt}) correspond to regions with barriers $\E$
before and after a reconfiguration event, respectively.
Conservation of probability is ensured by $\omega(t)$,
which is the total rate of reconfiguration events at time~$t$.

Remarkably, the coupled equations (\ref{e:dpbydt})
and (\ref{e:omega_t}) are {\em identical}
to the trap model of Bouchaud, which is known to qualitatively
reproduce many features of spin glasses and supercooled
liquids~\cite{trap_weak,trap_tree,trap_int,trap_jphysa}.
Thus the model we have derived can also be viewed
as Bouchaud's trap model,
with a mapping from the energy barrier $\E$ to density $\rho$
that is reached via the two-stage process of first assuming that $\E$
is proportional
to the smallest region that can rearrange independently of its environment,
{\em \`a la} Adam and Gibbs, and then using the Chicago group's
free volume argument to relate the size of this region to its density.
The relationship with Bouchaud's trap model is useful as it
allows known analytical results to be transferred to this application,
as described in the following section.

\section{Comparison to the experiments}
\label{s:exp}

In this section we compare the predictions of the model
to the experimental results given in \cite{exp1,exp2,exp3,exp4}.
The general procedure employed throughout was to numerically integrate
$P(\E,t)$ in time according to the master equation~(\ref{e:master})
from an initial state~$P(\E,0)$,
using the method described in Appendix~\ref{s:numerics}.
Ideally $P(\E,0)$ would be chosen to mimic
the distribution of density in the apparatus after the preparation phase,
but since such information is not available we have instead employed the
natural choice of $P(\E,0)=\mu(\E)$, which formally corresponds
to an instantaneous `quench' from $\T=\infty$.
No significant deviations are expected for other initial conditions
after an initial transient.
Once $P(\E,0)$ was fixed, the constant $A$ in (\ref{e:mapping}) was chosen
by trial-and-error to give an initial density close to
the experimental value $\rho(0)\approx0.58$.
The density $\rho(t)$ was extracted
at regular intervals by numerical evaluation of~(\ref{e:rhofrome}).

Each simulation was repeated for two different choices of~$\mu(\E)$,
namely an exponential $\mu(\E)=\frac{1}{\E_{0}}{\rm e}^{-\E/\E_{0}}$
and a Gaussian
$\mu(\E)=\sqrt{2/\pi\sigma^{2}}{\rm e}^{-\E^{2}/2\sigma^{2}}$,
where without loss of generality we now choose units such that
$E_{\rm 0}=\sigma=1$.
Other $\mu(\E)$ were also considered for the compaction under constant $\T$
described in Sec.~\ref{s:constgamma}
and were found to give the same behaviour for $t\lta10^{4}$ taps,
indicating that the model is robust to the particular choice of $\mu(\E)$
over the experimental timeframe.
However, this robustness does {\em not} extend to the
$t\rightarrow\infty$ limit, where it is already known that
different $\mu(\E)$ can give qualitatively different behaviour.
This is discussed thoroughly in~\cite{trap_jphysa}, but in brief,
an exponential tail $\mu(\E)\sim{\rm e}^{-\E}$ gives rise
to a {\em glass transition} at $\T=1$\,, in the sense that
an equilibrium solution only exists for $\T>1$\,.
This can be seen by simply
setting $\partial P/\partial t=0$ in the master equation~(\ref{e:master}),

\begin{equation}
P_{\rm eqm}(\E)\equiv\lim_{t\rightarrow\infty}P(\E,t)
=\omega(\infty)\,{\rm e}^{\E/\T}\mu(\E)\:\:,
\label{e:eqm}
\end{equation}

\noindent{}which is not normalisable for $\T\leq1$
if $\mu(\E)\sim{\rm e}^{-\E}$,
and hence equilibrium cannot be reached.
By contrast, if $\mu(\E)$ decays more rapidly than exponentially,
{\em e.g.} if it has a Gaussian tail, then an equilibrium
solution exists for all $\T>0$, although the equilibration time
may be excessively large for small~$\T$.
Note that this model quite generally predicts that the
limiting density $\rho_{\infty}=\lim_{t\rightarrow\infty}\rho(t)$
is a monotonic decreasing function of~$\T$.
A proof of this is given in Appendix~\ref{s:eqm_rho}.

\subsection{Constant excitation intensity}
\label{s:constgamma}

\begin{figure}
\centerline{\psfig{file=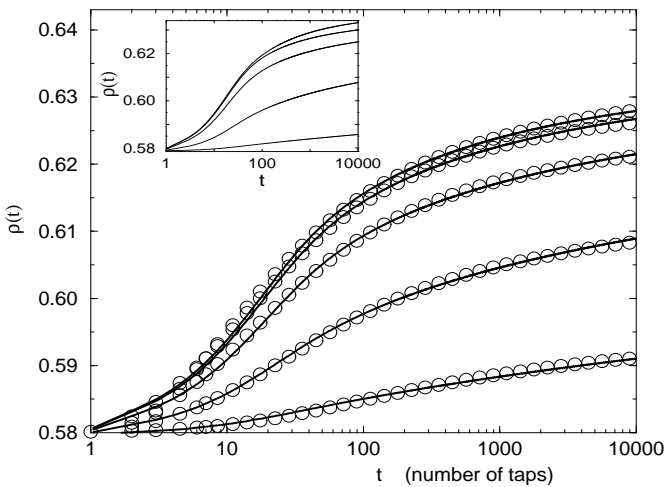,width=3.8in}}
\caption{Plot of $\rho(t)$ versus $t$ on log-linear axes for a Gaussian
$\mu(\E)=\sqrt{2/\pi}{\rm e}^{-\E^{2}/2}$ with $\rho_{\rm max}=0.64$,
$\omega_{0}=0.1$ and $A=0.053$.
From bottom to top, the lines correspond to
$\T=0.005$, 0.03, 0.1, 0.2 and 0.3 respectively.
The circles are fits to the empirical law~(\ref{e:logrelax}).
{\em (Inset)} The corresponding results for an
exponential $\mu(\E)={\rm e}^{-\E}$ and $A=0.05$, with
$\T=0.002$, 0.02, 0.1, 0.2 and 0.5\,.}
\label{f:logrelax}
\end{figure}

Simulation results for the mean density $\rho(t)$
over a range of $\T$ is given in Fig.~\ref{f:logrelax}.
Also given are fits to the empirical law~(\ref{e:logrelax}),
demonstrating that it is well obeyed
with either an exponential and Gaussian $\mu(\E)$.
We have also checked and found similar logarithmic behaviour
for a selection of other $\mu(\E)$, such as
uniform on $[\E_{0},\E_{1}]$, both with $\E_{0}=0$ and~$\E_{0}>0$,
Gaussian with a non-zero mean, Cauchy,
and exponential limited to the range $[\E_{0},\E_{1}]$.
However, logarithmic relaxation is {\em not} expected for
pathological $\mu(\E)$ such as $\delta(\E-\E_{0})$ or
$\exp(-{\rm e}^{\E})$.

The logarithmic behaviour can be understood by considering
the scaling solution to the master equation (\ref{e:master})
already found by Monthus and Bouchaud~\cite{trap_jphysa}.
They demonstrated that, after a short
transient, $P(\E,t)$ can be expressed in terms of a
single scaling variable $u$ as

\begin{equation}
P(\E,t)=\frac{1}{\T}\,u\,\phi(u)\:,\hspace{0.3in}
u=\frac{{\rm e}^{\E/\T}}{\omega_{0}\,t}\:\:.
\label{e:scaling}
\end{equation}

\noindent{}Strictly speaking this is only true for an exponential $\mu(\E)$
below the glass point, but it was also demonstrated that a Gaussian $\mu(\E)$
admits a similar scaling solution
until a time $t^{*}\sim\omega_{0}^{-1}{\rm exp}(1/\T^{2})$,
which may lie well beyond the experimental timeframe when $\T$ is small.
The physical picture underlying this scaling behaviour
is that the sizes of the cooperatively rearranging regions,
which are proportional to~$\E=\T\ln(\omega_{0}tu)$, are increasing
logarithmically in time.
A logarithmic increase in domain size has also been found in the
Tetris model~\cite{Mario_logdomain}.

Over timescales for which the scaling solution~(\ref{e:scaling}) holds,
the density can be expressed in terms of $\phi(u)$ by changing variables
from $\E$ to $u$ in~(\ref{e:rhofrome}),

\begin{equation}
\frac{\rho(t)}{\rho_{\rm max}}=
1-\int^{\infty}_{\frac{1}{\omega_{0}t}}\:
\frac{\phi(u)}{1+\displaystyle\frac{\T}{A\rho_{\rm max}}\ln(\omega_{0}tu)}
\,{\rm d}u\:\:.
\label{e:scalrho}
\end{equation}

\noindent{}The similarity of this expression
to the empirical law~(\ref{e:logrelax}) is striking.
The primary difference is that here we must integrate over
a distribution of~$u$,
which will in general introduce corrections to the simple logarithmic law.
The simulation results in Fig.~\ref{f:logrelax} demonstrate that any such
corrections are at most small.

The form of the theoretical prediction (\ref{e:scalrho}) makes it
difficult to calculate the fitting parameters $\Delta\rho$, $B$ and $\tau$
in the empirical law (\ref{e:logrelax}).
However, one parameter that can trivially be fixed is
the projected final density~$\rho_{\rm f}$\,,
which is always equal to $\rho_{\rm max}$ here,
regardless of $\T(\Gamma)$.
In contrast, the experiments seem to indicate that $\rho_{\rm f}$
is a non-monotonic function of $\Gamma$~\cite{exp1}.
There is no easy way to resolve this discrepancy.
For instance, one cannot simply assume that~$\rho_{\rm max}$
is itself a function of~$\Gamma$,
{\em i.e.} \mbox{$\rho_{\rm max}=\rho_{\rm max}(\Gamma)$}.
Quite apart from the conceptual difficulties this would invoke for the
physical meaning of $\rho_{\rm max}$\,,
it would allow situations in which {\em negative} free volume could arise,
for instance by first
allowing a system to relax arbitrarily close to
$\rho_{\rm max}(\Gamma)$ and then suddenly changing to
a $\Gamma'$ for which
$\rho_{\rm max}(\Gamma')<\rho_{\rm max}(\Gamma)$.
By definition, $\bar{v_{\rm f}}$ would then be negative.
Note that this contradiction is not specific to this model but will
arise whenever the definition of free volume (\ref{e:vfbar})
is used.

\begin{figure}
\centerline{\psfig{file=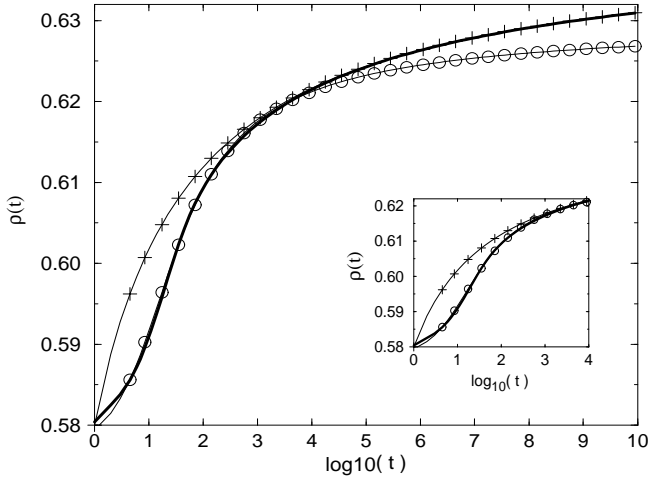,width=3.8in}}
\caption{Comparison between different choices of fitting parameters
over different ranges of~$t$.
The thick line is the density $\rho(t)$ for a Gaussian $\mu(\E)$
at $\T=0.1$ with $\rho_{\rm max}=0.64$\,.
The circles correspond to the fit
$\rho(t)=0.63-0.053/(1+2.3\ln(1+t/16))$
and the crosses correspond to
$\rho(t)=\rho_{\rm max}-0.06/(1+0.24\ln t)$.
{\em (Inset)} The same plots over the experimental timeframe
$1\leq t\leq10^{4}$.}
\label{f:large_t}
\end{figure}

\begin{figure}
\centerline{\psfig{file=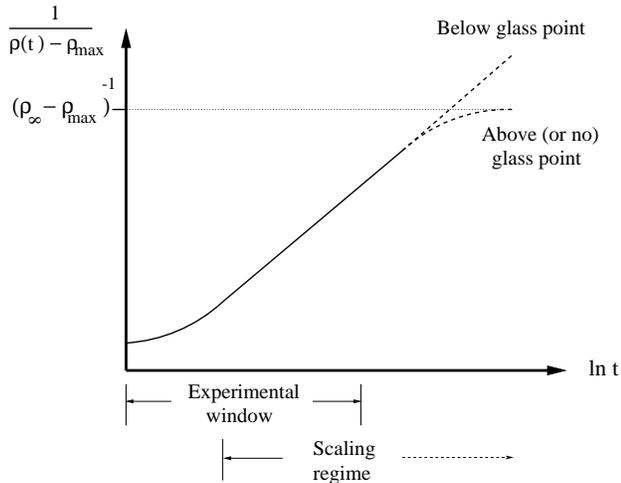,width=3.2in}}
\caption{Schematic of densification under constant~$\T$.
The line becomes straight, indicating that the density is
logarithmically increasing towards~$\rho_{\rm max}$\,,
once the system enters into the scaling regime,
which overlaps with the experimental window.
The scaling regime continues indefinitely if 
$\mu(\E)\sim{\rm e}^{-\E/\T_{\rm g}}$
and $\T\leq\T_{\rm g}$\,.
For $\T>\T_{\rm g}$\,, or for a $\mu(\E)$ with a tail that decays faster
than exponentially, the scaling behaviour ceases at some
late time and the density reaches its equilibrium density
$\rho_{\infty}<\rho_{\rm max}$\,.}
\label{f:schematic}
\end{figure}

We believe the solution to this problem lies in the range of $t$
over which the data fitting has been performed.
As mentioned previously, the scaling solution~(\ref{e:scaling}),
and hence the logarithmic relaxation, only applies
after a short transient, typically
$t\stackrel{>}{\scriptstyle\sim}10^{2}-10^{3}$ taps.
However, we have found that it is still possible to attain a
very reasonable fit to the empirical law
over the whole range $0\leq t\leq10^{4}$, but
{\em only at the expense of predicting the wrong $\rho_{\rm f}$}\,.
This is clearly demonstrated in Fig.~\ref{f:large_t},
which shows that a fit that works well for
$0\leq t\leq10^{4}$ fails when extrapolated to larger~$t$,
whereas fixing $\rho_{\rm f}=\rho_{\rm max}$ gives an initially
poorer fit but recovers the correct asymptotic behaviour.
Transferring this insight to the experiments
suggests that discarding the first 1-10\%
of the experimental data points and then repeating the fitting procedure
would result in a similar logarithmic compaction law as before,
but with $\rho_{\rm f}$ independent of~$\Gamma$.
The various time regimes in this model are
summarised schematically in Fig.~\ref{f:schematic}.

\subsection{Annealing curve}
\label{s:anneal}

Further insight into the nature of the system's relaxation properties
can be gained by allowing the tap intensity to vary in time,
which roughly corresponds to varying the temperature
in other slowly relaxing systems~\cite{Hammann,StAndrews}.
Two time-dependencies will be considered in this paper.
The first is the `annealing curve,' which was experimentally attained
by cyclically ramping $\Gamma$ in a stepwise fashion between
some high value  $\Gamma=\Gamma_{1}$ and $\Gamma=0$~\cite{exp2,exp3}.
Slowly decreasing $\Gamma$ removes low density local
configurations without creating many new ones,
hence the term `annealing.'
The second protocol for varying $\Gamma$ will be investigated in the
next subsection.

\begin{figure}
\centerline{\psfig{file=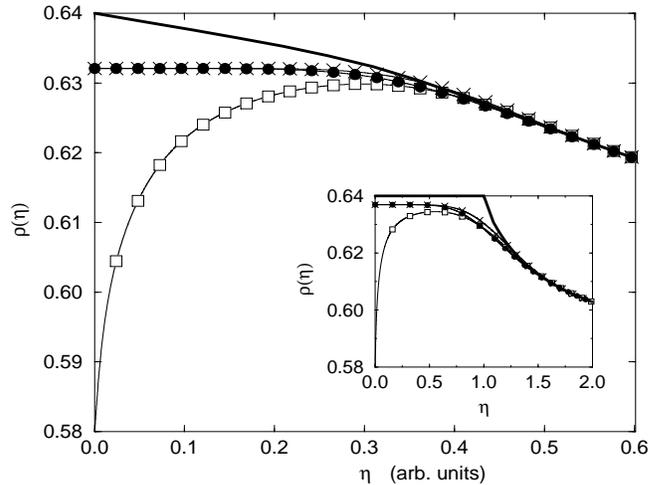,width=3.8in}}
\caption{The annealing curve for a Gaussian $\mu(\E)$.
The symbols refer to the initial increase in $\T$ up to $\T=0.6$
(open squares),
the decrease to $\T=0$ (filled circles) and the second increase (crosses).
For each leg, $\T$ was varied smoothly over $t_{\rm leg}=10^{6}$ taps.
The thick line is the equilibrium density.
{\em (Inset)} The same for an exponential $\mu(\E)$ over a range
of $\T$ that includes the glass point.
}
\label{f:anneal}
\end{figure}

\begin{figure}
\centerline{\psfig{file=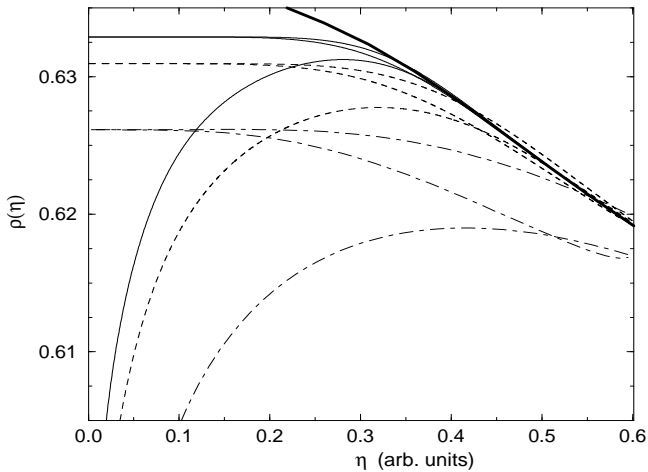,width=3.8in}}
\caption{Variation of the annealing curves with cooling rate
for a Gaussian $\mu(\E)$.
Results are given for a total time per leg of $t_{\rm leg}=10^{7}$
(thin solid line),
$10^{5}$ taps (dashed line) and $10^{3}$ taps (dot-dashed line).
The second and third legs are reversible for
$t_{\rm leg}\stackrel{>}{\scriptstyle\sim}10^{5}$.
The thick line is the equilibrium density.
}
\label{f:var_anneal}
\end{figure}

The annealing curve for this model is obtained by
allowing $\T$ to smoothly vary from 0 to some value $\T_{1}$
to 0 to $\T_{1}$ again,
where the duration of each leg is denoted by~$t_{\rm leg}$\,.
Simulation results for $t_{\rm leg}=10^{6}$
are given in Fig.~\ref{f:anneal}.
The experimental annealing curve has a similar shape, except
that the initial density increase for small $\Gamma$ is noticeably slower
than that for small~$\T$~\cite{exp2}.
This may simply be due to a non-trivial mapping from $\Gamma$ to~$\T$,
as discussed in Sec.~\ref{s:interp}.
Note that the second and third legs in Fig.~\ref{f:anneal}
form a reversible curve which is
nonetheless out of equilibrium for small~$\T$.
Observe also the presence of a narrow hysteresis loop,
which is
also present in microscopic models~\cite{tetris_aging,Prados,Sinai}
but has never been systematically searched for in experiments.
The area of this hysteresis loop decreases for slower cooling rates,
as demonstrated in Fig.~\ref{f:var_anneal}.

To interpret these results in a glassy context,
recall that the initial conditions were chosen to conform to
the equilibrium state at \mbox{$\T=\infty$}, {\em i.e.}
$P(\E,0)=P_{\rm eqm}(\E)|_{\T=\infty}=\mu(\E)$.
This would be valid if the initial low density configuration in the
experiments corresponded to an equilibrium state for very
large tapping intensity~$\Gamma$,
which seems plausible.
Thus the start of the first leg corresponds to
a rapid {\em quench} from high $\T$ to $\T\approx0$,
leaving the system far from equilibrium.
The rate of compaction is initially rapid but slows as the density,
and hence the relaxation times, increase.
For sufficiently high $\T$, the density reaches and starts to follow
the equilibrium curve, rapidly erasing the memory of its history.
As $\T$ is lowered a second time, this time
corresponding to a slow quench,
the system remains near equilibrium until some value
$\T_{0}$ (which depends on the cooling rate)
when the relaxation time rapidly increases 
and the system essentially freezes.
Thus the difference between the first and third legs in Fig.~\ref{f:anneal}
can be understood
as the recovery from a rapid and slow quench, respectively.

\subsection{Shift in the excitation intensity}
\label{s:tempcyc}

\begin{figure}
\centerline{\psfig{file=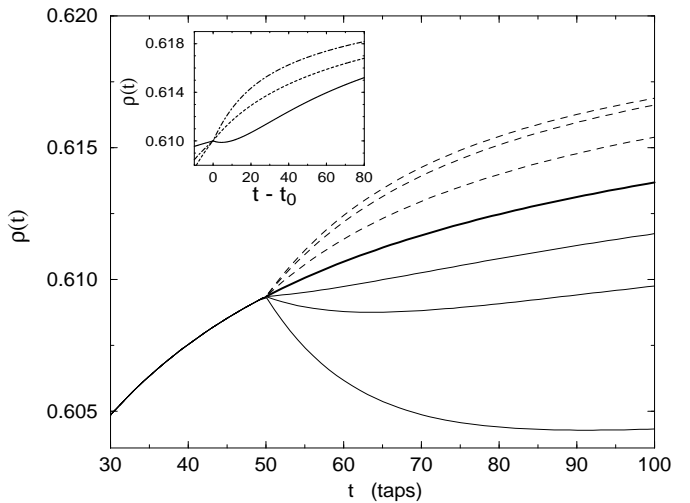,width=3.8in}}
\caption{Response to a shift from $\T=0.5$ to $\T=0.5+\Delta\T$ at
a time $t=50$ for a Gaussian $\mu(\E)$.
From top to bottom, the lines refer to
$\Delta\T=-0.3$, -0.2 and -0.1 (dashed lines), 0 (thick line)
and 0.1, 0.2 and 0.5 (thin solid lines).
{\em (Inset)} Here $\T=0.1$ (solid line), 0.3 (dashed line)
or 0.5 (dot-dashed line) until the first time $t_{0}$ when
$\rho(t_{0})\geq0.61$, afterwhich $\T$ is fixed at 0.3 in each case.
}
\label{f:pitchfork}
\end{figure}

Recent experiments have investigated the effect of allowing
$\Gamma$ to `shift' from a constant value $\Gamma_{0}$ to another
constant value $\Gamma_{1}$ at a given time~$t_{0}$~\cite{exp4}.
It was found that the system evolved in a way that depended
on its history as well as its current density and excitation
intensity~$\Gamma$, representing a form of memory akin to
that in glassy systems~\cite{Hammann,StAndrews}.
Plotted in Fig.~\ref{f:pitchfork} are the corresponding results for this
model, where $\T=\T_{0}=0.5$ until $t=t_{0}=50$ taps,
when it changes to $\T_{1}=\T_{0}+\Delta\T$.
The sign of the initial density change is opposite to the sign of $\Delta\T$,
as in the experiments, although this is not entirely general and
the behaviour is reversed if $t_{0}$ is too small.
Also shown in the inset is the case when $\T$ is changed from
different $\T_{0}$ to the same value $\T_{1}=0.3$ when the density
reaches a predetermined value.
Again there is qualitative agreement with the experiments.

\begin{figure}
\centerline{\psfig{file=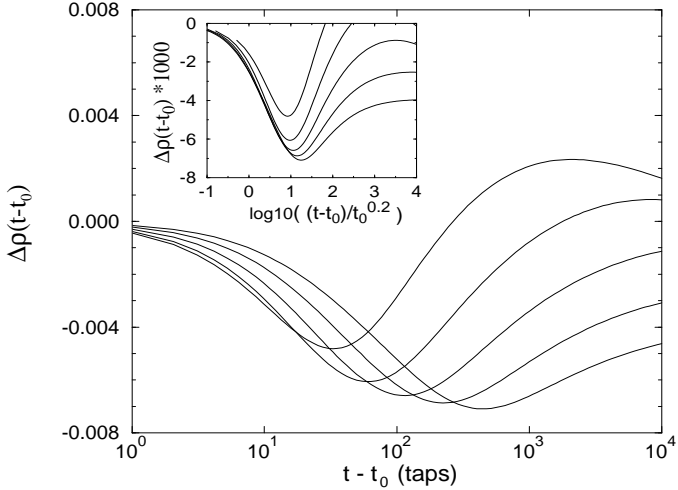,width=3.8in}}
\caption{Response to a shift from $\T=0.1$ to $\T=0.5$ at a time $t_{0}$
for a Gaussian~$\mu(\E)$, where
$\Delta\rho(t-t_{0})$ is the density difference between perturbed and
unperturbed systems.
From top the bottom on the right hand side, the lines refer to
$t_{0}=10^{3}$, $10^{4}$, $10^{5}$, $10^{6}$ and $10^{7}$ respectively.
An exponential $\mu(\E)$ behaves similarly.
{\em (Inset)} The same data plotted against $(t-t_{0})/t_{0}^{\alpha}$,
where $\alpha=0.1/0.5$ is the ratio of $\T$ before and after the shift.
}
\label{f:response}
\end{figure}

The analogy with glass systems suggests that the timescale of
the response to a shift in $\T$ at $t_{0}$ should scale with $t_{0}$
in some manner~\cite{tshift_sg}.
With this insight, we now make the following prediction,
in the hope it may be tested experimentally.
Let $\Delta\rho(t-t_{0})$ be the difference in density at time $t$
between the perturbed system and an unperturbed one,
{\em i.e.} one with $\Delta\T=0$\,.
Plotted in Fig.~\ref{f:response} is $\Delta\rho(t-t_{0})$ for a shift from a
low to a high $\T$ at times $t_{0}=10^{3}$, $10^{4}$, $10^{5}$,
$10^{6}$ and $10^{7}$.
In each case there is a well-defined time for the peak response
$t^{\rm resp}$, which increases with $t_{0}$\,.
Known results for the trap model with an exponential $\mu(\E)$
suggest that $t^{\rm resp}\sim t_{0}^{\T_{0}/\T_{1}}$,
where the exponent is independent of $t_{0}$~\cite{tshift_trap}.
We find this to be a good first approximation to our data,
as demonstrated by the inset to Fig.~\ref{f:response},
although there are corrections arising from the
non-linear mapping from $\E$ to~$\rho$, which distorts the
underlying scaling behaviour in~$\E$.
There are also additional small corrections when
using a Gaussian $\mu(\E)$.

\begin{figure}
\centerline{\psfig{file=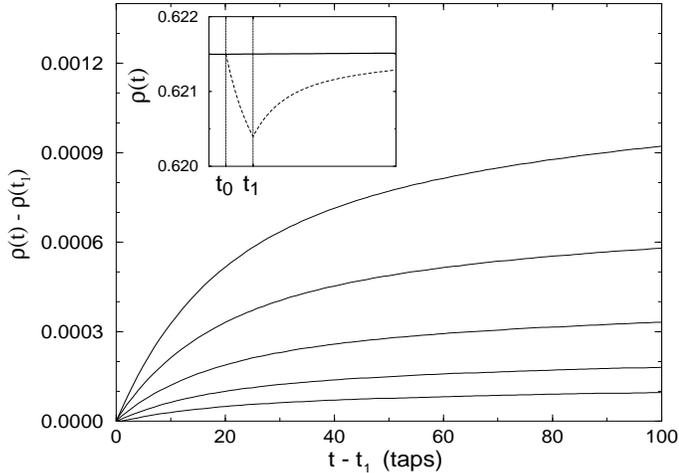,width=3.8in}}
\caption{Plot of the recovery from a short time at a higher $\T$
for a Gaussian $\mu(\E)$.
Here the system has been relaxed for a time
$t_{0}=10^{4}$ taps at $\T=0.1$, then held at $\T=0.3$ until
$t_{1}=t_{0}+\delta t$ when it reverts back to $\T=0.1$ again.
From bottom to top, the lines refer to $\delta t=1$, 2, 4,
8 and 16, respectively.
{\em (Inset)} The raw data for $\delta t=16$ (dashed line)
compared to the unperturbed system (solid line).
}
\label{f:recovery}
\end{figure}

Finally, the experiments briefly investigated what happens
when $\Gamma$ is allowed to return to its initial value after
$\delta t$ taps at a higher value~$\Gamma_{1}$~\cite{exp4}.
For comparison, the equivalent results from this model are given in
Fig.~\ref{f:recovery}.
The observed trend is in accord with the experimental observations.
A full study of this variation in $\T(t)$ for all
$\T_{0}$\,, $\T_{1}$\,, $t_{0}$ and $\delta t$ is beyond
the scope of this paper and will not be discussed further here.

\subsection{Fluctuations and power spectra}
\label{s:powspec}

In a finite system the density in equilibrium is not constant
but fluctuates about its mean value.
To investigate density fluctuations in this model,
a different version of the code was employed which explicitly
simulates a system consisting of $N$ separate subsystems
(details given in Appendix~\ref{s:numerics}).
Fig.~\ref{f:denfluct} shows the probability distribution $Q(\Delta\rho)$
of fluctuations $\Delta\rho\equiv\rho(t)-\rho_{\rm modal}$
for $N=500$ and different~$\T$.
To first approximation $Q(\Delta\rho)$ is Gaussian,
but it is slightly skewed towards lower densities,
becoming more so as $\T$ is lowered.
The skewness arises from the non-linear mapping from
$\E$ to $\rho$, which exaggerates fluctuations to lower densities
whilst suppressing those to high densities.
There is also a cut-off for very large $|\,\Delta\rho\,|$\,,
when $P(\E,t)$ has deviated significantly from $P_{\rm eqm}(\E)$.
The experiments exhibited Gaussian fluctuations with some
anomalous deviations for $\Delta\rho>0$~\cite{exp2};
this is discussed below.

\begin{figure}
\centerline{\psfig{file=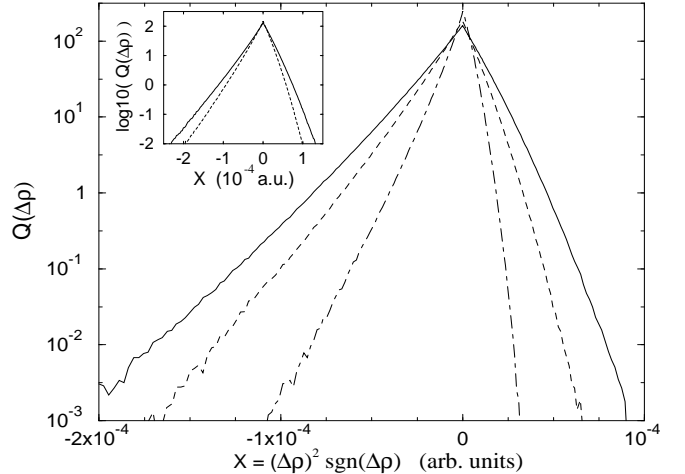,width=3.8in}}
\caption{Fluctuations around the modal density $\rho_{\rm modal}$
in equilibrium for an $N=500$ element system
with a Gaussian $\mu(\E)$.
$Q(\Delta\rho)$, the probability of a fluctuation
$\Delta\rho=\rho(t)-\rho_{\rm modal}$\,,
is plotted against $X=(\Delta\rho)^{2}{\rm sgn}(\Delta \rho)$ on a log-linear
plot, which would give a symmetrical triangle if $Q(\Delta\rho)$
was Gaussian.
The different lines refer to
$\T=1$ (solid), $\T=0.8$ (dashed) and $\T=0.6$ (dot-dashed).
{\em (Inset)} The same for an exponential $\mu(\E)$ with
$\T=2$ (solid line) and $\T=1.6$ (dashed line).
}
\label{f:denfluct}
\end{figure}

\begin{figure}
\centerline{\psfig{file=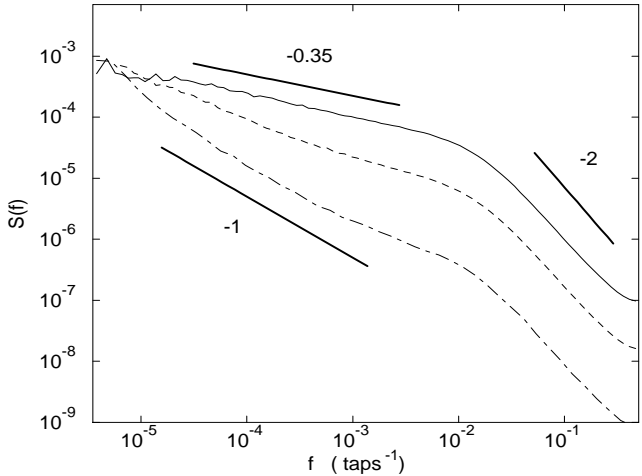,width=3.8in}}
\caption{
$S(f)$, the power spectrum of frequency~$f$,
for an $N=500$ system and a Gaussian $\mu(\E)$,
at $\T=0.6$ (solid line), $\T=0.4$ (dashed line) and
$\T=0.2$ (dot-dashed line, also vertically shifted by
a factor of 500 for clarity).
Each $S(f)$ was calculated over $\approx10^{8}$ points.
The slopes of the thick line segments are indicated.
To speed convergence, the initial configuration was chosen as
the known equilibrium state $P_{\rm eqm}(\E)$
for $N=\infty$~(\ref{e:eqm}),
although the $\T=0.2$ system was still evolving towards
its slightly different finite $N$ equilibrium state during the run.
This accounts for the anomalous steep slope for small~$f$.
}
\label{f:ps_gauss}
\end{figure}

The power spectra $S(f)$ of density fluctuations in equilibrium
for various $\T$ and a Gaussian $\mu(\E)$ is given in
Fig.~\ref{f:ps_gauss}.
$S(f)\sim1/f^{2}$ for $f$ greater than a high frequency shoulder $f_{\rm H}$\,,
where $f_{\rm H}$ is only weakly dependent on~$\T$,
although it should be stressed that this model focuses on cooperative
relaxation modes and is not intended to describe the
high frequency, single particle dynamics.
For low frequencies, $S(f)$ appears to obey non-trivial power law behaviour
$S(f)\sim1/f^{\delta}$, where the exponent $\delta$
can be {\em very approximately}
fitted to $\delta\approx1-\T$ over the range $10^{-5}<f<10^{-3}$.
However, the analysis given in Appendix~\ref{s:spec_anal}
shows that this is not the true asymptotic behaviour and
$S(f)\rightarrow1/f^{0}$ as $f\rightarrow0$.
The crossover to $1/f^{0}$ behaviour occurs around a low frequency shoulder
$f_{\rm L}$\,, where $f_{\rm L}\rightarrow0$ rapidly as $\T\rightarrow0$.
For an exponential $\mu(\E)$ there is only one shoulder frequency
separating the high frequency, $1/f^{2}$ regime from a low
frequency regime in which $S(f)\sim1/f^{\delta}$, where
$\delta=2-\T$ for $1<\T<2$ and $\delta=0$ for $\T\geq2$.
Note that there is no $1/f^{0}$ region for $\T<2$, even though the system
is in equilibrium.
This apparent anomaly is explained in Appendix~\ref{s:spec_anal}.

In the experiments, the power spectra were found to obey
non-trivial power law behaviour $S(f)\sim1/f^{\delta}$,
with $\delta=0.9\pm0.2$, between two corner frequencies
$f_{\rm L}$ and $f_{\rm H}$ that both decreased as $\Gamma$
was lowered~\cite{exp2}.
More complex behaviour was observed for larger~$\Gamma$
towards the bottom of the apparatus.
The results from this model are in partial agreement;
for instance, it is still one of few that can exhibit a $1/f^{\delta}$
regime with $\delta\approx1$ (see also~\cite{BS}).
There are some discrepancies, but these may simply be due to processes not
currently incorporated into the model, such as single particle
dynamics or the existence of metastable, high-density crystalline domains.

\begin{figure}
\centerline{\psfig{file=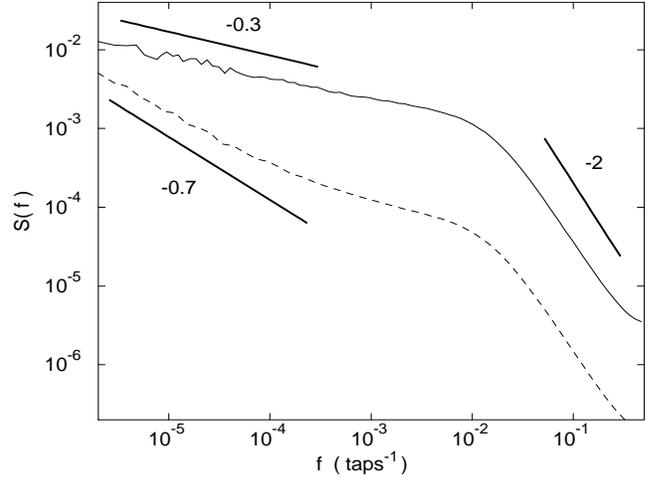,width=3.8in}}
\caption{$S(f)$ in equilibrium for an exponential $\mu(\E)$ and
$N=500$, with $\T=1.7$ (solid line, also shifted vertically
by a factor of 10) and $\T=1.3$ (dashed line).
The analytical predictions from Appendix~\ref{s:spec_anal}
are indicated by the thick line segments.
}
\label{f:ps_exp}
\end{figure}

\section{Discussion of the model parameters}
\label{s:interp}

Given the success of this model in reproducing
the experimental phenomenology, it is natural to ask
if its principle assumptions can be placed on a firmer foundation.
In particular, a number of parameters introduced in Sec.~\ref{s:freevol}
have so far been treated somewhat heuristically.
To redress the balance, we now discuss the physical
interpretation of some of these parameters.
A more thorough analysis may be possible by detailed comparison
with a microscopic model, for instance.

\subsubsection{The noise parameter $\T(\Gamma)$}

It was stressed during the derivation of this model that the noise
parameter $\T$ need not bear any relation to
the concept of granular temperature~\cite{temp_Ogawa,temp_X1}.
By the same token, the use of the term `equilibrium' to describe
the statistical steady state
merely refers to a {\em dynamic} equilibrium, without supposing
any analogy with a {\em thermodynamic} one.
Instead, $\T$ was defined in the broadest sense of simply
giving some measure of the degree of excitation of the system during
a single tap.
This loose definition makes finding the
precise relationship with $\Gamma$ difficult.
Nonetheless it is still possible to predict the overall shape of~$\T(\Gamma)$,
as we now argue.

For $\T$ to be non-zero, the particles must at the very least separate
from their nearest neighbours.
Experiments on vibrated granular systems often claim to find
some critical $\Gamma_{\rm c}$ such that the
relative motion of the particles is either minimal
or non-existent for $\Gamma<\Gamma_{\rm c}$
(usually $1\lta\Gamma_{\rm c}\lta2$, see {\em e.g.}~\cite{rev1,exp1,mri}\,).
A facile explanation for this is to suppose that a granular body
is held together by frictional forces, and that relative motion between
adjacent particles is not possible until some static friction
threshold has been overcome.
Since all the normal contact forces are proportional to~$g$,
the distribution of threshold forces will also scale with~$g$ and
thus the relevant parameter would indeed be $\Gamma=a_{\rm max}/g$.
However, friction is not the only relevant mechanism.
For instance, particle separation will still occur in
vertical one dimensional columns, where friction clearly plays no part.
Even in this case theory suggests that the relevant
parameter is again~$\Gamma$,
at least in the limit of hard spheres~\cite{1dtheory}.

Assuming that a well-defined $\Gamma_{\rm c}$ exists,
the overall shape of $\T(\Gamma)$ will obey
$\T(\Gamma)\approx0$ for small~$\Gamma$,
only significantly deviating from zero for
$\Gamma\gta\Gamma_{\rm c}$\,.
Just this qualitative behaviour has been found in
simulations of horizontally vibrated systems~\cite{horiz}.
As a corollary, the annealing curves presented in Figs.~\ref{f:anneal}
and~\ref{f:var_anneal} will be flatter for small $\Gamma$ when
plotted against $\Gamma$ rather than~$\T$, in better agreement with
the experimental graphs~\cite{exp2,exp3}.

\subsubsection{The prior distribution $\mu(\E)$}

The Gaussian and exponential $\mu(\E)$ employed in the simulations
were chosen as plausible first guesses of the real $\mu(\E)$.
To calculate the actual $\mu(\E)$ is a non-trivial problem, but a
first step might be to re-express $\mu(\E)$ in terms of $\psi(\freevol)$,
the distribution of free volume after reconfiguration.
Using $\E=z^{*}\Delta\E=Av_{\rm g}/\freevol$~(\ref{e:minsize}),
this gives

\begin{equation}
\mu(\E)=\frac{Av_{\rm g}}{\E^{2}}
\,\psi\left(\frac{Av_{\rm g}}{\E}\right)\:\:.
\label{e:psivf}
\end{equation}

\noindent{}In principle, $\psi(\freevol)$ could be found from a
microscopic model, such as the parking lot
model~\cite{exp2,Prados,parkinglot,Talbot}, for which the
free volume is also the void volume.

Note that, from~(\ref{e:psivf}),
the tail of $\mu(\E)$, which is so important to the
long-time relaxational properties of Bouchaud's trap model,
can be related to the $\freevol\rightarrow0^{+}$ behaviour of $\psi(\freevol)$.
For example, if $\psi(\freevol)$ vanishes according to
$\psi(\freevol)\sim\exp(-\alpha/\freevol)$,
then $\mu(\E)$ will have an exponential tail and the trap model predicts
a glass transition at a finite noise intensity
\mbox{$\T=Av_{\rm g}/\alpha$}.
Similarly, $\psi(\freevol)\sim\exp(-\alpha/\freevol^{2})$
corresponds to a $\mu(\E)$ with a Gaussian tail.

\subsubsection{The constant $A=a\Delta\E$}
\label{s:regimes}

Even though $A$ has
been treated as an arbitrary constant and fixed by the initial
conditions, its component factors $a$ and $\Delta\E$
have a physical interpretation, as we now discuss.
In (\ref{e:minsize}), $a$ is defined as the constant of proportionality
between~$z^{*}$, the smallest number of particles that can
cooperatively rearrange, and the ratio $v_{\rm g}/\freevol$\,.
In the Chicago group's original argument~\cite{exp2}, $a$ was set to~1;
however, we prefer not to fix $a$ at any particular value
and suggest that it may depend upon particle properties
such as their shape.
For instance, highly irregular particles will obstruct motion more
effectively than rounder particles of the same volume, and so should
have a higher value of~$a$.

$\Delta\E$ was originally defined as a gravitational potential energy barrier.
Indeed, assuming that the particles interact via hard core repulsion,
this is the {\em only} available potential energy scale in the system.
This suggests that $\Delta\E$ is proportional to the mean
vertical displacement between adjacent particles.
If so, then our implicit assumption that $\Delta\E$ is independent
of $\T$ and $\rho$ is compatible with
the hard sphere Monte Carlo simulations
of Barker and Mehta~\cite{Barker_MC}, who demonstrated that the
distribution of contact angles between particles is roughly
constant over a wide range of shaking amplitudes.

More importantly, if $\Delta\E$ is gravitational potential
energy, then inspection of the reconfiguration rate (\ref{e:micrate})
indicates that $\T$ must also have units of energy, with an energy
scale that is presumably coupled to the driving.
For definiteness, suppose $\T$ can be written as $\T=mgA_{0}f(\Gamma)$,
where $m$ is the typical mass of the particles,
$A_{0}$ is the amplitude of the driving and $f(\Gamma)$ is some function
of the dimensionless parameter~$\Gamma=a_{\rm max}/g$
(possibly with a threshold around $\Gamma\approx\Gamma_{\rm c}$
as discussed earlier).
Similarly, write $\Delta\E\sim mgr$, where $r$ is the typical particle radius.
Since $\Delta\E$ and $\T$ only appear in the ratio~$\Delta\E/\T$,
$m$ and $g$ will cancel and the dynamics of the model will
depend on {\em two} dimensionless quantities,
namely $\Gamma$ and a dimensionless {\em displacement}~$A_{0}/r$.
The existence of a second relevant dimensionless parameter implies
that the behaviour in response to high
amplitude, low frequency driving may be qualitatively different from a low
amplitude, high frequency driving with the same value of~$\Gamma$.
This possibility has not yet been explored in the experiments,
which seem to have focused on the low frequency regime.

Note that we could equally have expressed $\T$ in terms of the
kinetic energy supplied by the driving,
{\em i.e.} $\T=mv_{0}^{2}\,\tilde{f}(\Gamma)$,
where $v_{0}$ is the typical driving velocity.
However, this is not an independent energy scale
as $v_{0}$ can be dimensionally related to $a_{\rm max}$ and
$A_{0}$ by $v_{0}\sim\sqrt{a_{\rm max}A_{0}}$\,.
We only mention this latter alternative because simulations often show
scaling plots in terms of $\Gamma$ and $v_{0}$
(see {\em e.g.}~\cite{horiz} and references therein).

\section{Summary and conclusions}
\label{s:summ}

To summarise, we have constructed a simple model for weakly excited granular
media that combines the Chicago group's free volume argument with elements
of the supercooled liquid theory of Adam and Gibbs.
Integration of the master equation has shown that the model
behaves in a similar manner to the
experiments for each of the situations considered.
Some slight discrepancies remain with the power spectra,
but these may be due to mechanisms currently lacking from the model,
such as ordering effects and crystallinity, depth dependency or wall effects.
It would be interesting to see if any of these mechanisms could be
incorporated into an extended version of model.
It may also be possible to introduce orientational degrees of freedom
and compare the results to recent experiments on nylon rods~\cite{exp_rods}.

The model has also been used to predict the manner in
which the time of the peak response to a shift in $\Gamma$ at $t=t_{0}$
scales with~$t_{0}$\,, as discussed in Sec.~\ref{s:tempcyc}.
This prediction could be tested experimentally and may help
to differentiate between the large number of models that have so far
been proposed
\cite{tetris_prl,tetris_full,tetris_aging,Mario_logdomain,Mario_FDT,Prados,parkinglot,Talbot,Gavrilov,BS,Sinai,Grinev,Linz,Barker},
as it seems unlikely that they will all give the same scaling behaviour.
Further insight into the physical mechanisms underlying the compaction
process could be gained by measuring the typical size of reconfiguring
regions as a function of time,
or by seeing if the locations of such regions are spatiotemporally correlated.
Such measurements could be performed in simulations, or by direct
visualisation of two dimensional experiments~\cite{Warr2D},
for instance.

Finally, we note that the relationship between granular media and
glasses can be given a more intuitive appeal by the following simple argument.
Consider sand poured from a great height into a container.
When the particles first hit the surface of the forming sandpile, the
direction in which they bounce will essentially be random,
giving rise to a large random velocity component.
This corresponds to a highly excited state
with (in our notation) a high~$\T$.
However, the particles will rapidly lose their kinetic energy by inelastic
collisions and will soon come to rest, jamming under gravity into
a static, disordered configuration with $\T=0$.
It is not difficult to see how this sequence of events can be related
to the rapid `quench' of a supercooled liquid or other glass-forming
material.

Just after the initial submission of this work, we became aware
of a master equation for the glass transition due to Dyre~\cite{dyre},
which is similar to Bouchaud's equation studied in this paper
but with a built-in cut-off in the range of allowed energies.
Also, it has been brought to our attention that
the two regimes of vibration mentioned in Sec.~\ref{s:regimes}
have previously been discussed in the context of size segregation by
Mehta and Barker~\cite{new_mehta}.

\section*{Acknowledgements}
\label{s:ack}

The author would like to thank Mike Cates for helpful discussions
and careful reading of the manuscript, and also Joachim Wittmer,
Mario Nicodemi, Alan Bray, Jean-Philippe Bouchaud,
Robin Stinchcombe and Suzanne Fielding for
stimulating discussions on the experiments and this model.
We would also like to thank Jeppe Dyre for bringing our attention
to reference~\cite{dyre}, and Anita Mehta and Gary Barker for
reference~\cite{new_mehta}.
This work was funded by UK EPSRC grant no. GR/M09674.

\appendix

\section{Simulation details}
\label{s:numerics}

The bulk of the simulation results were obtained by numerical integration
of the continuous master equation~(\ref{e:master}).
$P(\E,t)$ was defined on a mesh of
points $P_{ij}=P(i\,\delta\E,j\,\delta t)$,
where \mbox{$0\leq i\leq i_{\rm max}$} and \mbox{$j\geq0$}.
Care was taken to ensure that $E_{\rm max}\equiv i_{\rm max}\,\delta\E$
was set sufficiently high that there was no significant cut-off to $P(\E,t)$
for large~$\E$.
To iterate over a single time step~$\delta t$,
$\omega(t)$ was found from numerical integration of (\ref{e:omega_t})
and then assumed to remain constant over the required time interval.
This allowed the time evolution equation (\ref{e:dpbydt})
to be solved and
$P_{ij+1}$ found from $P_{ij}$ $\forall i$.
The whole distribution was then renormalised by a factor
$(\sum_{i}\,P_{ij})^{-1}$
to correct for the non-conservation of probability resulting from the
assumption of a constant $\omega(t)$.
For relaxation under constant~$\T$, simulation times were
improved by employing a geometric mesh with a linearly
increasing time step \mbox{$\delta t\propto t$}.
This allowed for times up to $t=10^{10}$ to be reached
with only modest CPU time.

For the density fluctuations investigated in Sec.~\ref{s:powspec},
the continuous master equation was of no use and an alternative
method was employed which explicitly included finite size effects.
This involved assigning $N$ array elements a barrier $E_{\rm i}$\,,
\mbox{$i=1\ldots N$}, according to the chosen initial conditions.
At every time step $\delta t=1$, each element was assigned a new
barrier with probability $\omega_{0}\,{\rm e}^{-\E_{i}/\T}$,
where the new barrier values were drawn from the prior~$\mu(\E)$.
The density $\rho_{i}$ of each element was found by inverting
the mapping~(\ref{e:mapping}),
and the mean density calculated by straightforward summation,
$\rho(t)=\frac{1}{N}\sum\rho_{i}$\,.

\section{Monotonicity of $\rho_{\infty}(\T)$ on $\T$}
\label{s:eqm_rho}

In this appendix it is shown that the asymptotic density
$\rho_{\infty}=\lim_{t\rightarrow\infty}\rho(t)$
is a monotonic decreasing function of~$\T$
for essentially any $\mu(\E)$.
If an equilibrium state exists, it takes the form
$P_{\rm eqm}(\E)=\omega_{\infty}(\T)\,{\rm e}^{\E/\T}\mu(\E)$
and hence from~(\ref{e:rhofrome}),

\begin{eqnarray}
\frac{\rho_{\infty}(\T)}{\rho_{\rm max}} &=&
1-\int_{0}^{\infty}\frac{\omega_{\infty}(\T){\rm e}^{\E/\T}\mu(\E)}
{1+\E/\A\rho_{\rm max}}\,{\rm d}\E\:\:,
\label{appb:main}\\
\omega_{\infty}(\T)&\equiv&\lim_{t\rightarrow\infty}\omega(\T,t)
=\left(\int_{0}^{\infty}{\rm e}^{\E'/\T}\mu(\E')\,{\rm d}\E'\right)^{-1}.
\end{eqnarray}

\noindent{}Differentiating (\ref{appb:main})
with respect to $\T$ and rearranging gives

\begin{eqnarray}
\lefteqn{
\frac{\T^{2}}{\omega^{2}_{\infty}(\T)\,\displaystyle\rho_{\rm max}}
\,\frac{\partial\rho_{\infty}(\T)}{\partial\T}
=}\nonumber\\
&&\int_{0}^{\infty}\int_{0}^{\infty}
(\E-\E')
\frac{\mu(\E)\mu(\E'){\rm e}^{(\E+\E')/\T}}{1+E/A\rho_{\rm max}}
\,{\rm d}\E\,{\rm d}\E'\:\:.
\label{e:app_mid}
\end{eqnarray}

\noindent{}After the change of variables $u=\E+\E'$ and $v=\E-\E'$
and substituting $v\rightarrow-v$ over the domain $v<0$,
the right hand side of (\ref{e:app_mid}) transforms to

\begin{eqnarray}
\lefteqn{
-\,\int_{u=0}^{\infty}\int_{v=0}^{u}\,
v^{2}\,{\rm e}^{u/\T}
\mu\left(\frac{u+v}{2}\right)\mu\left(\frac{u-v}{2}\right)
}\nonumber\\
&&\times
\Big[
(2A\rho_{\rm max}+u+v)(2A\rho_{\rm max}+u-v)\Big]^{-1}
\,{\rm d}u\,{\rm d}v\:\:.
\label{e:app_finint}
\end{eqnarray}

Since all the factors inside the integral (\ref{e:app_finint})
are positive, it can be trivially deduced that

\begin{equation}
\frac{\partial\rho_{\rm max}}{\partial\T}\leq0
\hspace{0.2in}\mbox{for all $\T$.}
\end{equation}

\begin{figure}
\centerline{\psfig{file=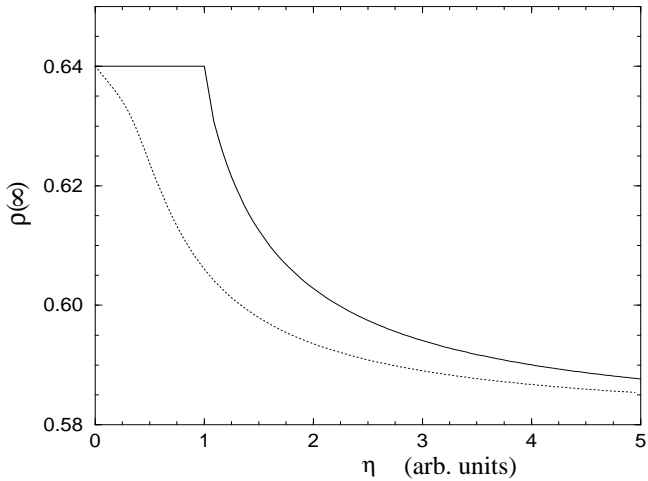,width=3.8in}}
\caption{Plot of $\rho_{\infty}\equiv\lim_{t\rightarrow\infty}\rho(t)$
against $\T$ for an exponential $\mu(\E)={\rm e}^{-\E}$ (solid line)
and the Gaussian $\mu(\E)=\sqrt{2/\pi}{\rm e}^{-\E^{2}/2}$
(dotted line) for $\rho_{\rm max}=0.64$.
Note that the plateau for $\T\leq1$ in the exponential case
corresponds to the out-of-equilibrium situation where all the moments
of $P(\E,t)$ diverge as $t\rightarrow\infty$.
}
\label{f:eqmrho}
\end{figure}

\noindent{}Equality is attained in only two cases.
The first is if the integrand
is strictly zero over the entire range~$v>0$,
which can only happen in the trivial case of a single valued distribution
$\mu(\E)=\delta(\E-\E_{0})$.
The second and more important situation 
is over a range of $\T$ for which no equilibrium state exists.
In this case, all the moments of $P(\E,t)$
diverge as \mbox{$t\rightarrow\infty$} and
$\rho(t)\rightarrow\rho_{\rm max}$ from~(\ref{e:mapping}).
A plot of $\rho_{\infty}(\T)$ versus $\T$
for an exponential and a Gaussian~$\mu(\E)$
is given in Fig.~\ref{f:eqmrho}.

\section{Analysis of the power spectra}
\label{s:spec_anal}

The small $f$ behaviour of the power spectra of density fluctuations
$S(f)$ is analytically derived in this appendix,
which extends the range of the numerical observations
discussed in Sec.~\ref{s:powspec}.
The time-dependent spectrum near the glass point
has already been derived in~\cite{trap_tree};
here we consider the $f\rightarrow0^{+}$ limit in
equilibrium for general $\mu(\E)$ over a wider range of $\T$.

Since each local region is assumed to relax independently of its
environment, the total power spectrum $S(f)$ is just the spectrum for a
single region with a relaxation time~$\tau$
averaged over $\Phi(\tau)$,
the distribution of relaxation times in equilibrium,

\begin{equation}
S(f)\propto\int\,
\frac{\tau\,\Phi(\tau)}{1+(2\pi f\tau)^{2}}\,
{\rm d}\tau\:\:.
\label{e:sf}
\end{equation}

\noindent{}The small $f$ behaviour of (\ref{e:sf}) depends on
the asymptotic behaviour of $\Phi(\tau)$ for large~$\tau$.
If $\Phi(\tau)$ decays faster than $\tau^{-2}$, then the \mbox{$f\equiv0$}
limit exists and $S(f)$ exhibits the expected
$1/f^{0}$ noise for low frequencies.
However, if $\Phi(\tau)\sim\tau^{-x}$ with $1<x<2$,
then $S(f)\sim 1/f^{2-x}$, as can be readily seen
by substituting for $f\tau$ in (\ref{e:sf})
(note that $x>1$ since $\Phi(\tau)$ is normalisable).

For the trap model, $\Phi(\tau)$ can be found for any given $\mu(\E)$
by simply making the change of
variables $\tau=\frac{1}{\omega_{0}}{\rm e}^{\E/\T}$ into the
expression for $P_{\rm eqm}(\E)$, equation~(\ref{e:eqm}).
Thus, an exponential $\mu(\E)\sim{\rm e}^{-E/\T_{\rm g}}$
gives $\Phi(\tau)\sim\tau^{-\T/\T_{\rm g}}$,
implying that $S(f)\sim1/f^{2-\T/\T_{\rm g}}$ for
\mbox{$\T_{\rm g}<\T<2\T_{\rm g}$}\,.
This confirms that $S(f)\not\rightarrow1/f^{0}$ for this range of~$\T$,
even though the system is in equilibrium.
The usual $1/f^{0}$ behaviour is recovered when \mbox{$\T\geq2\T_{\rm g}$}\,,
which also applies for all $\T$ when $\mu(\E)$ decays faster
than exponentially.
In particular, a Gaussian
$\mu(\E)\sim{\rm e}^{-\E^{2}/2\sigma^{2}}$
leads to an equilibrium distribution of relaxation times of the form

\begin{equation}
\Phi(\tau)\sim\tau^{-\frac{\T^{2}}{2\sigma^{2}}\ln(\omega_{0}\tau)}\:\:,
\end{equation}

\noindent{}which is suggestive of a power law with a slowly varying
exponent $x=\frac{\T^{2}}{2\sigma^{2}}\ln(\omega_{0}\tau)$.
Thus one would expect $S(f)$ to exhibit approximate power law
behaviour over a wide range of $f$,
reverting to $1/f^{0}$ only for frequencies comparable to the
`largest' relaxation time $\omega_{0}\tau^{*}\sim{\rm e}^{\sigma^{2}/\T^{2}}$.
Any attempt to fit $S(f)$ to a power law will give an exponent
that depends on the range of $f$ considered as well as the ratio $\T/\sigma$.


\end{multicols}

\end{document}